\documentclass[12pt]{iopart}
\usepackage{iopams}

\begin{document}

\title{Spectrum of a duality-twisted Ising quantum chain}

\author{Uwe Grimm}

\address{Applied Mathematics Department, 
Faculty of Mathematics and Computing,
The Open University, Walton Hall, 
Milton Keynes MK7 6AA, UK}

\begin{abstract}
The Ising quantum chain with a peculiar twisted boundary condition is
considered. This boundary condition, first introduced in the framework
of the spin-$1/2$ XXZ Heisenberg quantum chain, is related to the
duality transformation, which becomes a symmetry of the model at the
critical point. Thus, at the critical point, the Ising quantum chain
with the duality-twisted boundary is translationally invariant,
similar as in the case of the usual periodic or antiperiodic boundary
conditions. The complete energy spectrum of the Ising quantum chain is
calculated analytically for finite systems, and the conformal
properties of the scaling limit are investigated. This provides an
explicit example of a conformal twisted boundary condition and a
corresponding generalised twisted partition function.
\end{abstract}



Recently, there has been a lot of interest in generalised twisted
partition functions of conformal field theory \cite{PZ1,PZ2,PZ3,CS}
and the corresponding conformal twisted boundary conditions in
solvable lattice models of statistical mechanics \cite{CMOP}. Examples
of such boundary conditions leading to the appearance of ``exotic''
spinor operators in the corresponding conformal field theory had
previously been considered for quantum spin chains, in particular for
the Ising and $3$-state Potts quantum chains \cite{GS,S}. In these
cases, the boundary conditions were derived from a mapping
\cite{AGR,ABGR2,PS,GR1,GR2,GS} between these models with $N$ sites and
the spin-$1/2$ XXZ Heisenberg model with twisted boundary conditions,
which for an even number of sites $2N$ yields the ``usual'' twisted
boundary conditions \cite{AGR,PS}, such as antiperiodic boundary
conditions for the Ising quantum chain, and results in rather exotic
boundary terms when considered for an odd number of sites $2N-1$
\cite{GS}.  The corresponding boundary conditions still allow for
translational symmetry, but only at the critical point. The
corresponding symmetry for the Ising and Potts models was identified
as duality, which maps the Ising Hamiltonian of the ordered phase onto
that of the disordered phase and vice versa. It becomes a proper
symmetry at the critical point, and the corresponding boundary terms
and generalised translation and parity operators were analysed in
\cite{GS,S} by means of an algebraic approach \cite{L}.

The conformal partition function corresponding to this duality twisted
boundary condition was derived in \cite{GS} from that of the twisted
XXZ Heisenberg spin chain \cite{ABGR1}. However, the spectrum of the
duality-twisted Ising quantum chain itself has never been calculated
explicitly. In fact, using the standard free-fermion approach
\cite{LSM}, it is possible to obtain the complete spectrum even for
finite chains.  This is the purpose of this Letter. The calculation
proceeds along the same path as for Ising quantum chains with
``generalised defects'' \cite{G}, as the boundary terms are actually
rather similar. However, the defects considered in \cite{G}, albeit
featuring rather general couplings of the last and the first spin in
the chain, did not include the duality-twisted boundary considered
here. For comparison, we treat the usual periodic and antiperiodic
boundary conditions in the same way.

Consider an Ising quantum spin chain consisting of $N$ sites, so the
Hilbert space is $\mathcal{H}_{N}=(\mathbb{C}^{2})^{\otimes
N}\cong\mathbb{C}^{2^N}$. We introduce local spin operators, acting at
site $j$, by
\begin{equation}
\sigma^{w}_{j} = \left(\bigotimes_{k=1}^{j-1}\mathbb{I}\right)
\otimes \sigma^{w}_{}\otimes 
\left(\bigotimes_{k=j+1}^{N}\mathbb{I}\right),
\end{equation}
where $w$ stands for $x$, $y$ and $z$, and $\mathbb{I}$ denotes the
unit operator. In the canonical basis, the spin operators
$\sigma^{w}_{}$ are represented by the Pauli matrices
\begin{equation}
\sigma^{x}_{}=\pmatrix{0&1\cr 1&0},\qquad
\sigma^{y}_{}=\pmatrix{0&-\rmi\cr\rmi&0},\qquad
\sigma^{z}_{}=\pmatrix{1&0\cr 0&-1},
\end{equation}
and $\mathbb{I}$ is the $2\times2$ unit matrix.

The critical Ising quantum chain with periodic $(+)$ and antiperiodic
$(-)$ boundary conditions is defined by the Hamiltonians
\begin{equation}
H^{(\pm)} = -\frac{1}{2}\left(\sum_{j=1}^{N-1}
\sigma^{x}_{j}\sigma^{x}_{j+1}
+\sum_{j=1}^{N}\sigma^{z}_{j}\pm\sigma^{x}_{N}\sigma^{x}_{1}\right).
\label{IsingPA}
\end{equation}
The Ising quantum chain is related to the transfer matrix of the
classical zero-field square-lattice Ising model at the critical
temperature through an anisotropic limit. The normalisation factor is
chosen such that the fermion velocity in the scaling limit is equal to
unity. The duality-twisted Ising quantum chain is given by \cite{GS,S}
\begin{equation}
H^{(\mathrm{d}\pm)} = -\frac{1}{2}\left(\sum_{j=1}^{N-1}
\sigma^{x}_{j}\sigma^{x}_{j+1}+
\sum_{j=1}^{N-1}\sigma^{z}_{j}
\pm\sigma^{y}_{N}\sigma^{x}_{1}\right).
\label{IsingD}
\end{equation}
Note that, besides the coupling between $\sigma^{y}_{N}$ and
$\sigma^{x}_{1}$, the Hamiltonians \eref{IsingD} and \eref{IsingPA}
also differ in the diagonal terms, as there is no operator
$\sigma^{z}_{N}$ in \eref{IsingD}. This is also the reason why such
boundary terms were not considered as a ``generalised defect'' in
\cite{G}. Because the generalised parity transformation for the
duality-twisted Hamiltonian maps the two signs at the boundary onto
each other \cite{S}, and since the two Hamiltonians are related by
complex conjugation, they have the same spectrum, and the momenta of
the eigenstates differ by a sign.

The Hamiltonians $H^{(\pm)}$ and $H^{(\mathrm{d}\pm)}$ commute with
the operator
\begin{equation}
Q = (\sigma_{}^{z})^{\otimes N} = \prod_{j=1}^{N}\sigma_{j}^{z},
\end{equation}
so they have a global $C_{2}$ symmetry. We denote the corresponding
projectors by
\begin{equation}
P_{\pm}^{} = \frac{1}{2}\left(\mathbb{I}_{N}\pm Q\right),
\end{equation}
where $\mathbb{I}_{N}=\mathbb{I}^{\otimes N}$ is the identity operator
on $\mathcal{H}_{N}$.  In addition, they commute with properly defined
translation operators, which take the boundary condition into account,
see \cite{L,GS,S} for details.

In order to obtain local boundary conditions in the fermionic
language, we play the usual trick and pass to ``mixed-sector''
Hamiltonians \cite{BCS} defined as
\begin{eqnarray}
\tilde{H}^{(\pm)} & = & H^{(\pm)}P_{+}^{} +  H^{(\mp)}P_{-}^{},
\label{pams}\\
\tilde{H}^{(\mathrm{d}\pm)} & = & H^{(\mathrm{d}\pm)}P_{+}^{} +  
H^{(\mathrm{d}\mp)}P_{-}^{}.
\label{dms}
\end{eqnarray}
Performing a Jordan-Wigner transformation \cite{LSM}
\begin{equation}
c_{k}^{} = \left(\prod_{j=1}^{k-1}
        \sigma_{j}^{z}\right)\sigma_{k}^{-},
\qquad
c_{k}^{\dagger} = \left(\prod_{j=1}^{k-1} 
        \sigma_{j}^{z}\right) \sigma_{k}^{+},
\end{equation}
we rewrite $\tilde{H}^{(\pm)}$ and $\tilde{H}^{(\mathrm{d}\pm)}$ as
bilinear expressions in the fermionic creation and annihilation
operators $c_{k}^{\dagger}$ and $c_{k}^{}$,
\begin{equation}
\tilde{H} = 
\frac{N}{2}\mathbb{I}_{N} + \sum_{j,k=1}^{N} 
\left[ D_{j,k} c_{j}^{\dagger}c_{k}^{}
+ \frac{1}{2} \left( E_{j,k}^{}
c_{j}^{\dagger}c_{k}^{\dagger}
+ E_{k,j}^{\ast}
c_{j}^{}c_{k}^{} \right) \right],
\label{quad}
\end{equation}
where the $N\times N$ matrices $D$ and $E$ depend on the boundary
conditions
\begin{eqnarray}
2D_{j,k}^{\pm} &=&-2\delta_{j,k}+\delta_{j+1,k}(1-\delta_{j,N})
+\delta_{j,k+1}(1-\delta_{k,N})\nonumber\\
& & \mp\delta_{j,1}\delta_{k,N}\mp\delta_{j,N}\delta_{k,1},\\
2E_{j,k}^{\pm} &=&\delta_{j+1,k}(1-\delta_{j,N})
-\delta_{j,k+1}(1-\delta_{k,N})\pm
\delta_{j,1}\delta_{k,N}\mp\delta_{j,N}\delta_{k,1},\\
2D_{j,k}^{\mathrm{d}\pm} &=&-2\delta_{j,k}(1-\delta_{j,N})+
\delta_{j+1,k}(1-\delta_{j,N})
+\delta_{j,k+1}(1-\delta_{k,N})\nonumber\\
& & \mp\rmi\,
\delta_{j,1}\delta_{k,N}\pm\rmi\,\delta_{j,N}\delta_{k,1},\\
2E_{j,k}^{\mathrm{d}\pm} &=&\delta_{j+1,k}(1-\delta_{j,N})
-\delta_{j,k+1}(1-\delta_{k,N})\mp\rmi\,
\delta_{j,1}\delta_{k,N}\pm\rmi\,\delta_{j,N}\delta_{k,1},
\end{eqnarray}
and $\ast$ denotes complex conjugation.  The quadratic form
\eref{quad} can be diagonalised to
\begin{equation}
\tilde{H} = \sum_{k=0}^{N-1}\Lambda_{k}^{}\eta_{k}^{\dagger}\eta_{k}^{} + 
e\,\mathbb{I}_{N}
\label{diag}
\end{equation}
by means of a Bogoliubov-Valatin transformation. Here,
$\eta_{k}^{\dagger}$ and $\eta_{k}^{}$ are new fermionic creation and
annihilation operators, $e$ is a constant, and the squared fermion
energies $\Lambda_{k}^2$ are given by the eigenvalues of the $2N\times
2N$ matrix
\begin{equation}
A=\pmatrix{D&E\cr -E^{\ast}&-D^{\ast}}^2
= \pmatrix{D^2-EE^{\ast}&(D-D^{\ast})E\cr
-(D-D^{\ast})E^{\ast}&(D^{\ast})^2-EE^{\ast}},
\end{equation}
see \cite{G} for details. The matrix $A$ has the property $CAC=A^{\ast}$,
where $C$ is the $2N\times 2N$ matrix with elements
\begin{equation}
C_{j,k} = \delta_{j+N,k}+\delta_{j,k+N}.
\end{equation}
An eigenvector $\Phi$ of $A$ corresponds to a proper
Bogoliubov-Valatin transformation if it satisfies
$C\phi=\phi^{\ast}$. As was shown in \cite{G}, the spectrum of $A$ is
doubly degenerate, and one can always choose the eigenvectors such
that they obey the conjugation condition. In this approach, we
effectively doubled the system, which is the reason why each fermion
energy occurs twice. We only need to consider ``half'' the eigenvalues
of $A$, thus removing this trivial degeneracy.  Furthermore, because
we can change the sign of the fermion energies $\Lambda_{k}$ in
\eref{diag} by a transformation that exchanges creation and
annihilation operators, resulting only in a different value of the
constant in \eref{diag}, we may choose all energies to be positive,
and ordered, so
$0\le\Lambda_{0}\le\Lambda_{1}\le\ldots\le\Lambda_{k-1}$. In this
case, the constant in \eref{diag} is just the corresponding
ground-state energy, $e=E_{0}$.

For the periodic and antiperiodic mixed-sector Hamiltonians
$\tilde{H}^{\pm}$ of \eref{pams}, it turns out that $A$ is in fact
block-diagonal, and both blocks are identical, because all matrix
elements are real.  So in this case we can limit ourselves to one
block immediately, and diagonalise the resulting $N\times N$
matrix. The solution of the eigenvalue equations
\begin{equation}
(D^2-EE^{\ast})^{(\pm)}\Psi_{k}^{(\pm)}=
(\Lambda_{k}^{(\pm)})^2\Psi_{k}^{(\pm)},
\end{equation}
with $k=0,1,\ldots,N-1$, are simply given by 
\begin{equation}
\Lambda_{k}^{(\pm)} = 2\sin(\case{p_{k}^{(\pm)}}{2}), \qquad 
(\Psi_{k}^{(\pm)})_{j} = \exp(\rmi\, p_{k}^{(\pm)}\! j),
\label{pasol}
\end{equation}
where the eigenvalues and the unnormalised eigenvectors are
parametrised by the momenta
\begin{equation}
p_{k}^{(+)} = \frac{(2k+1)\pi}{N},\qquad
p_{k}^{(-)} = \frac{2k\pi}{N}.
\end{equation}

For the duality-twisted mixed-sector Hamiltonians
$\tilde{H}^{(\mathrm{d}\pm)}$ \eref{dms}, the situation is more
complicated, because the matrix
$A^{(\mathrm{d})}:=A^{(\mathrm{d}+)}=A^{(\mathrm{d}-)}$ does not have
block form. Nevertheless, the bulk part of the equations, which does
not involve boundary terms, is still the same and can be solved as
above. The modified equations correspond to the rows $1$, $N-1$, $N$,
$N+1$, $2N-1$ and $2N$ of the matrix $A^{(\mathrm{d})}$. These yield
additional conditions, so one should look for a solution involving
terms $\exp(\pm\rmi\, p_{k}^{(\mathrm{d})} j)$ with site-dependent
coefficients at the boundary. Whereas it was not possible to solve the
equations in closed form for the generalised defects in \cite{G}, it
is possible in this case. One arrives at the solution
\begin{equation}
\Lambda_{k}^{(\mathrm{d})} = 
2|\sin(\case{p_{k}^{(\mathrm{d})}}{2})|,\qquad
p_{k}^{(\mathrm{d})} = \pm\frac{4k\pi}{2N-1},
\label{dsol}
\end{equation}
and the corresponding unnormalised eigenvectors of $A^{(\mathrm{d})}$
are
\begin{eqnarray}
(\Phi_{k}^{(\mathrm{d})})_{j}^{} = 
(\Phi_{k}^{(\mathrm{d})})_{j+N}^{\ast} & = &
\cos(\case{p_{k}^{(\mathrm{d})}}{2})\,
\exp(\rmi\, p_{k}^{(\mathrm{d})}\!j)\nonumber\\
&&+\rmi\,[\sin(\case{p_{k}^{(\mathrm{d})}}{2})-1]\,
\exp(-\rmi\, p_{k}^{(\mathrm{d})}\!j),\qquad
1\le j\le N\!-\!1,\nonumber\\
(\Phi_{k}^{(\mathrm{d})})_{N}^{}=
 (\Phi_{k}^{(\mathrm{d})})_{2N}^{\ast}&=&
1 - \sin(\case{p_{k}^{(\mathrm{d})}}{2}).
\label{dvec}
\end{eqnarray}
In general, positive and negative momenta in \eref{dsol} give two
linearly independent eigenvectors \eref{dvec} for the same eigenvalue.
However, there is only one vector for $k=0$, which corresponds to the
zero mode. A second, linearly independent eigenvector is
\begin{equation}
(\Phi_{0}^{(\mathrm{d})})_{j}^{}=\rmi\,(\delta_{j,N}-\delta_{j,2N}), 
\end{equation}
as can easily be checked.

The fermion excitation spectra \eref{pasol} and \eref{dsol} determine
the complete energy spectrum of the mixed-sector Hamiltonians
$\tilde{H}^{(\pm)}$ \eref{pams} and $\tilde{H}^{(\mathrm{d}\pm)}$
\eref{dms}, respectively, apart from the constant in \eref{diag},
which corresponds to the ground-state energy of the respective
Hamiltonian. This can be calculated by considering the trace of the
Hamiltonian. As
$\tr(\tilde{H}^{(\pm)})=\tr(\tilde{H}^{(\mathrm{d}\pm)})=0$, and
\begin{equation}
\tr\left(\sum_{k=0}^{N-1}\Lambda_{k}^{}\eta_{k}^{\dagger}\eta_{k}^{}
+E_{0}^{}\,\mathbb{I}_{N}^{}\right)
=N\left(\frac{1}{2}\sum_{k=0}^{N-1}\Lambda_{k}^{}+E_{0}^{}\right),
\end{equation}
the ground-state energies are given by
\begin{eqnarray}
-E_{0}^{(+)}=\sum_{k=0}^{N-1}
\sin(\case{(k+\frac{1}{2})\pi}{N})&=&
\frac{1}{\sin(\case{\pi}{2N})}=
\frac{2N}{\pi}+\frac{\pi}{12N}+\Or(N^{-3}),\label{eop}\\
-E_{0}^{(-)}=\sum_{k=0}^{N-1}
\sin(\case{k\pi}{N})&=&
\cot(\case{\pi}{2N})=
\frac{2N}{\pi}-\frac{\pi}{6N}+\Or(N^{-3}),\label{eom}\\
-E_{0}^{(\mathrm{d})}=\sum_{k=0}^{N-1}
\sin(\case{k\pi}{N\!-\!\frac{1}{2}})&=&
\frac{1+\cos(\case{\pi}{2N-1})}
{2\sin(\case{\pi}{2N-1})}\nonumber\\
&=&\frac{2(N-\frac{1}{2})}{\pi}-\frac{\pi}{24(N\!-\!\frac{1}{2})}+
\Or[(N\!-\!\case{1}{2})^{-3}]\nonumber\\
&=&\frac{2N}{\pi}-\frac{1}{\pi}-\frac{\pi}{24N}+\Or[N^{-2}]
\label{eod}
\end{eqnarray}
Comparing those with the finite-size corrections of the ground-state
energy expected from conformal field theory
\begin{equation}
-E_{0}=A_{0}N+\frac{\pi c}{6N}+\mathrm{o}(N^{-1}),
\label{fsc}
\end{equation}
we find that the ground-state energy per site in the thermodynamic
limit is given by
\begin{equation}
A_{0}=\lim_{N\rightarrow\infty}(-\frac{E_{0}^{+}}{N})=
\lim_{N\rightarrow\infty}(-\frac{E_{0}^{-}}{N})=
\lim_{N\rightarrow\infty}(-\frac{E_{0}^{\mathrm{d}}}{N})=
\frac{2}{\pi}
\end{equation}
and the central charge $c=\frac{1}{2}$ for the periodic chain. The
finite-size corrections for the other two cases result in ``effective
central charges'' $\tilde{c}^{-}=-1$ and
$\tilde{c}^{\mathrm{d}}=-1/4$, which correspond to excitations with
conformal dimension $x^{-}=(c-\tilde{c}^{-})/12=1/8$, the
magnetisation operator, and
$x^{\mathrm{d}}=(c-\tilde{c}^{\mathrm{d}})/12=1/16$, the ``exotic''
spinor operator with conformal spin $1/16$ \cite{GS}. Note that, for
the latter result, the result \eref{eod} shows that it is preferable
to define the scaling limit by using $N-\frac{1}{2}$ as the effective
system size in \eref{fsc}; for the naive scaling with $N$, a constant
surface term $A_1=1/\pi$ appears in \eref{fsc}, similar to what one
finds for free or fixed boundary conditions, or in the case of defect
lines \cite{G,BCS}.  In that sense, the duality-twisted chain acts as
a toroidal chain of $N-\frac{1}{2}$ sites rather than $N$ sites, in
accordance with the relation to the odd length spin-$1/2$ XXZ
Heisenberg spin chain \cite{GS}.

Using the appropriate scaling, the linearised low-energy spectrum
becomes
\begin{eqnarray}
\lim_{N\rightarrow\infty}[\frac{N}{2\pi}\,
(\tilde{H}^{+}-E_{0}^{+}\mathbb{I}_{N})]&=&
\sum_{r=0}^{\infty}[(r+\case{1}{2})a_{r}^{\dagger}a_{r}^{} +
(r+\case{1}{2})b_{r}^{\dagger}b_{r}^{}]\\
\lim_{N\rightarrow\infty}[\frac{N}{2\pi}\,
(\tilde{H}^{-}-E_{0}^{+}\mathbb{I}_{N})]&=&
\sum_{r=0}^{\infty}[ ra_{r}^{\dagger}a_{r}^{} +
(r+1)b_{r}^{\dagger}b_{r}^{}] + \frac{1}{8}
\end{eqnarray}
for the periodic and antiperiodic mixed-sector Hamiltonians, and
simply
\begin{equation}
\lim_{N\rightarrow\infty}[\frac{N\!-\!\frac{1}{2}}{2\pi}\,
(\tilde{H}^{\mathrm{d}\pm}-\frac{N\!-\!
\frac{1}{2}}{N}E_{0}^{+}\mathbb{I}_{N})]=
\sum_{k+0}^{\infty}[ra_{r}^{\dagger}a_{r}^{} +
(r+\case{1}{2})b_{r}^{\dagger}b_{r}^{}]+\frac{1}{16}
\end{equation}
for the duality-twisted case, where the fermionic operators $a_{r}$
and $b_{r}$ are obtained from the $\eta_{k}$ by renumbering. Taking
into account the momenta, and bosonising the low-energy theory in
terms of two generators $L_{0}$ and $\bar{L}_{0}$ of two commuting
Virasoro algebras with central charge $c=\frac{1}{2}$ by means of a
Sugawara construction, the first two yield the usual conformal torus
partition functions
$Z^{+}=(\chi_{0}+\chi_{1/2})(\bar{\chi}_{0}+\bar{\chi}_{1/2})$ and
$Z^{-}=2\chi_{1/16}\bar{\chi}_{1/16}$ for the Ising model, where
$\chi_{\Delta}$ denote the character function of the irreducible
representations of the $c=\frac{1}{2}$ Virasoro algebra with highest
weight $\Delta$. Depending on the sign of the boundary term, the last
corresponds to the combinations
$(\chi_{0}+\chi_{1/2})\bar{\chi}_{1/16}$ or
$\chi_{1/16}(\bar{\chi}_{0}+\bar{\chi}_{1/2})$ of characters, as
derived in \cite{GS} from the relation to the XXZ Heisenberg chain
with toroidal boundary conditions \cite{AGR,PS,GS}.

Hence, for this particular instance of conformally twisted boundary
conditions, the complete spectrum is known even for finite systems,
and the scaling limit can be performed explicitly. It would be
interesting to know whether the more general boundary conditions
introduced recently \cite{PZ1,PZ2,PZ3,CS,CMOP} can be interpreted in a
similar way, and what the corresponding symmetries are that allow the
definition of twisted toroidal boundary conditions.

\end{document}